\documentclass[fleqn,twoside]{article}
\usepackage{espcrc2}
\usepackage{graphicx}

\newcommand{\AmS}{{\protect\the\textfont2
  A\kern-.1667em\lower.5ex\hbox{M}\kern-.125emS}}

\title{Transversity signals at COMPASS}

\author{Federica  Sozzi\address{Trieste University and INFN Trieste, \\ 
         via Valerio, 2, 34127 Trieste, Italy\\
	federica.sozzi@ts.infn.it} (On behalf of the COMPASS collaboration)
}
       
\begin{document}

\begin{abstract}

COMPASS is a fixed target experiment  at  the CERN SPS, with a rich
 physics program focused on nucleon spin  structure  and on hadron
 spectroscopy.
 One of the main goals of the spin program is the measurement of the
transverse spin  distribution function $\Delta_T q(x)$ in semi-inclusive DIS off transversely polarized nucleons.
For this purpose  approximately 20\% of the running time in the years 2002 to
2004 with the longitudinally polarized muon beam of 160 GeV and with $^6$LiD polarized target 
 was used to collectdata with the  target polarized transversely with respect to the beam
direction.
The 2002 data have been already analysed and published. We present here the preliminary 
results from the full statistics for the Collins and Sivers single hadron  
asymmetries and for the transverse spin asymmetry
in hadron pair production.

\end{abstract}

\maketitle

\section{INTRODUCTION}
The nucleon spin structure can be described at leading twist by three parton
distribution functions~\cite{JaJi91}: the unpolarized distribution $q(x)$,  the  helicity 
distribution $\Delta q(x)$ and  the transversity distribution $\Delta_T q(x)$, 
 describing the probability in a transversely polarized nucleon 
of finding a quark with spin parallel to the nucleon spin.

The $\Delta_T q(x)$ distribution  is still largely unknown, and  this is the reason why  transversity nowadays  
has  an important role in  the scientific programme of different experiments:  
 HERMES at DESY,   RHIC at BNL and   COMPASS at CERN.  

The COMPASS collaboration is measuring transversity   in semi-inclusive DIS (SIDIS)  of leptons on transversely polarized nucleons, using different quark polarimeters:
azimuthal distribution of single  hadrons,
 azimuthal dependence of the plane containing hadron pairs,
and measurement of transverse polarization of baryons ($\Lambda$ hyperons).

In this contribution, we will present results on single hadron and hadron pair asymmetries.

\section{SINGLE HADRON ASYMMETRIES}

Due to its chiral odd
nature,  in SIDIS the $\Delta_T q(x)$ 
function can be assessed only together with 
 another chiral odd function.

One possibility is to measure it together with the so called Collins function $\Delta_T^0 D_q^h$~\cite{Collins:1993kk}.
According to Collins, the fragmentation of a transversely  polarized quark 
into an unpolarized hadron has an azimuthal dependence,
 with respect to the plane defined by the quark momentum and the quark spin.
 For this reason the event yield can be written as:
 \begin{equation}
N = N_0 \cdot (1+f \cdot P_t \cdot D_{nn} \cdot A_C \cdot \sin(\phi_C)) 
\label{eq:collfun}
\end{equation}
where $f$ is the target dilution factor, $P_t$ the target polarization and $ D_{nn}  = (1-y)/(1-y+y^2/2)$ the transverse spin
transfer coefficient from the initial to 
the struck quark. The Collins angle $\phi_C$ is defined as $\phi_h - \phi_{s'}$, where $\phi_h$ is the angle of the transverse momentum of the outgoing hadron and $\phi_{s'} = \pi - \phi_{s}  $ is the azimuthal angle of the struck quark spin ($\phi_{s}  $ is the azimuthal angle of quark  before the hard scattering). $A_C$ is the Collins asymmetry coming from the coupling between the  Collins fragmentation function and the transverse spin distribution:
\begin{equation}
A_C= \frac{\sum_q e_q^2 \, \Delta_T q(x) \, \Delta_T^0 D_q^h(z, p_T^2)}
       {\sum_q e_q^2 \,     q(x) \,       D_q^h(z, p_T^2)} 
\end{equation}
where  $z= E_h /(E_{l}-E_{l'})$ is the fraction of available energy carried by the hadron,
and $p_T$ is the hadron transverse momentum with respect to the virtual photon direction.

As is clear from eq.\ref{eq:collfun}, the Collins asymmetry  $A_C$ is  revealed as a $\sin \phi_{ C}$ modulation
in the number of produced hadrons.

Another azimuthal asymmetry is related to the Sivers effect~\cite{Sivers},
arising from a possible coupling 
of the intrinsic transverse momentum $k_T$ of unpolarized quarks to the spin of a transversely polarized nucleon. 
In this case the dependence of the number of  produced hadrons is:
 \begin{equation}
N = N_0 \cdot (1+f  \cdot P_t  \cdot A_S  \cdot \sin(\phi_S)) 
\end{equation}
where the Sivers angle $\phi_S$ is defined as $\phi_h - \phi_{s}$,
and the asymmetry $A_S$ probes the so called Sivers distribution function,  $\Delta_0^T q$:
\begin{equation}
A_{S}  =  \frac {\sum_q e_q^2 \, \Delta_0^T q (x,k_T) \, D^h_q  (z)}
{\sum_q e_q^2 \, q (x) \, D_q^h(z)} .
\end{equation}
In this case the asymmetry $A_S$ is revealed as a $\sin \phi_{ S}$ modulation 
in the number of produced hadrons. It has to be noted here that 
since the Collins and the Sivers angles are independent~\cite{colsiv}, 
it is possible to measure from the same data both the Collins and the Sivers effects.

\subsection{RESULTS}

We present the recent results for Collins and Sivers asymmetries 
on all the data available up to now.

The results from 
 the first physics run we had in 2002 have already been published~\cite{Alexakhin:2005iw}, while another paper containing all our statistics (2002-2004 data taking) is going to be published\cite{newart}.

The COMPASS experiment~\cite{compass}~\cite{FB03} uses a longitudinally polarized $\mu^+$ beam of 160 GeV/c and a 
two cells $^6$LiD target, whose nucleons are polarized transversely respect to the beam direction.  
The $^6$LiD material is characterized by a high dilution factor $f\sim$0.38, and the target polarization is about 50\%. 
The nucleons in the two target cells (upstream and downstream cell) have opposite polarizations
 to minimize systematic effects; the polarizations are reversed once per week.
The asymmetries are extracted using at the same time the information 
coming from both cells in two periods with opposite configuration; this is done  building the estimator:
\begin{equation}
A_j(\phi_{j}) = \frac{ N_{j,u}^{\uparrow}(\phi_{j})}
                     {N_{j,u}^{\downarrow}(\phi_{j})} \cdot
\frac{ N_{j,d}^{\uparrow}(\phi_{j})} 
                     {N_{j,d}^{\downarrow}(\phi_{j})} , \; \; \; \; \; j=C,S 
\label{eq:ratiop}
\end{equation}
and fitting the measured $A_j(\phi_{j})$ values with the functions
$p_0 \cdot(1. + A^{raw}_{C} \cdot \sin \phi_{C})$ and 
$p_0 \cdot(1. + A^{raw}_S \cdot \sin \phi_{S})$,
where $p_0$ is a free parameter. The  physical asymmetries are extracted from the raw asymmetries $A^{raw}_{C}$ and $A^{raw}_{S}$ 
knowing f, $P_T$ and also $D_{nn}$ in the Collins case.

In the analysis, the events are considered only 
if one primary vertex with incoming and outgoing $\mu$ is found in the target region.
To select DIS events, the photon virtuality
$Q^2$  is taken above 1 (GeV/c)$^2$, 
the fractional energy of the virtual photon $y$   between 0.1 and 0.9, and 
the invariant mass of the final hadronic
                             state $W$ above 5 GeV/c$^2$.

The hadron sample on which the asymmetries are computed consist of
 all the hadrons coming  from the reaction vertex with  $p_T >$ 0.1 GeV/c and
$z>$0.2.
Also a leading hadron sample is considered, using the most energetic hadron with $z > 0.25$ for each event.

\begin{figure}[htb]
\includegraphics[width=0.42\textwidth]{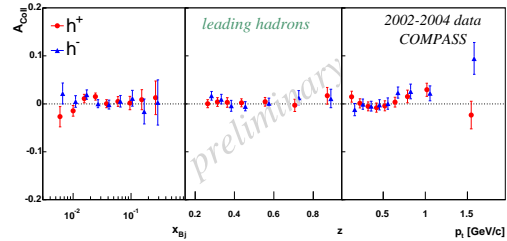}
\caption{Collins asymmetries for positive (circles) and negative (triangles) leading hadrons as a function of $x$, $z$ and $p_T$.}
\label{fig:coll_h}
\end{figure}

\begin{figure}[htb]
\includegraphics[width=0.41\textwidth]{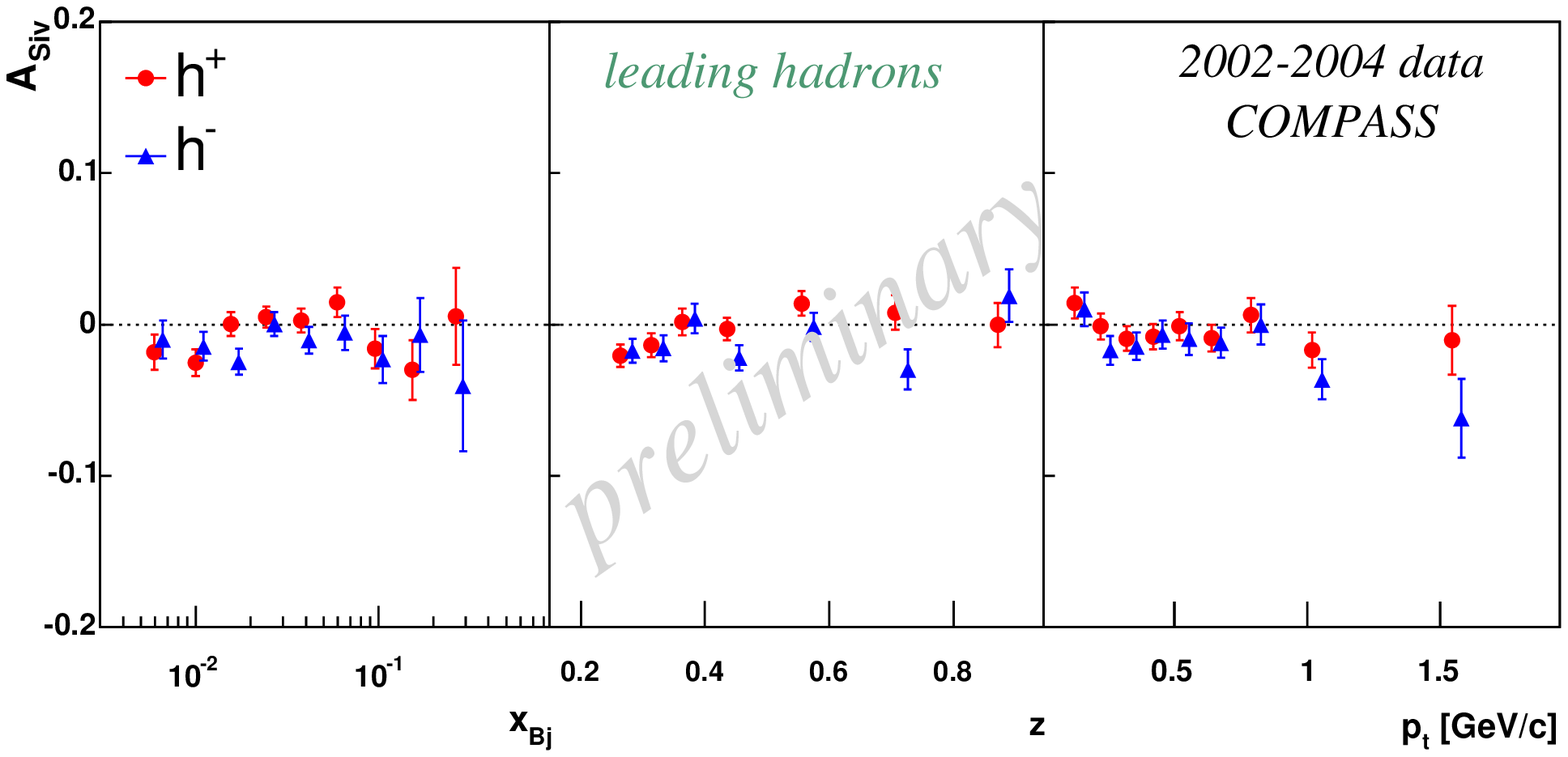}
\caption{Sivers asymmetries for positive (circles) and negative (triangles) leading hadrons as a function of $x$, $z$ and $p_T$.}
\label{fig:siv_h}
\end{figure}

The results on the Collins and Sivers asymmetries, plotted as a function of the kinematical variables $x$, $z$ and $p_t$, 
are shown in fig.~\ref{fig:coll_h} and ~\ref{fig:siv_h} for  leading hadrons.
A lot of work has been done in order to investigate possible sources of systematic effects; 
the conclusion of all
the studies is that the systematic error is considerably smaller than the statistical one, 
shown in the plots. 

All the asymmetries measured are small and compatible with zero within the quoted errors. The 
result is compatible with some phenomenological works  suggesting the cancellation between the p and n contribution in an isoscalar target as the one
used by COMPASS.

Results on identified hadrons have  recently 
been produced.
The hadron identification is done using the 
 RICH-1 detector~\cite{rich1} in the COMPASS spectrometer. 
 
Before applying RICH-1 in the analysis,
several studies have been performed in order to ensure the detector stability throught time:
the number of hadron identified as $\pi$  and $K$ normalized to the number of reconstructed tracks 
have been monitored in time. The runs or spills were rejected in
correspondance of a deviation larger than 3 times the standard deviation of the distributions.

The identification procedure relies on cuts on a likelihood function associated with each ring 
detected in the RICH-1 detector.
 The Likelihood function describes the signal number of photons taking into account the Frank and Tamm equation;
 the background number of photons, coming from other particles in the event and the beam halo, is evaluated from real data through the analysis of the  photon detector occupancy.

Since particle identification with RICH is possible only above the Cherenkov threshold, 
a lower limit on hadron momentum has been applied: it is $\sim$3 GeV/c for $\pi$ and $\sim$10 GeV/c for $K$, values a little above
the corresponding Cherenkov threshold in order to assure a minimum number of emitted photons. 
The upper limit for momentum has been chosen as 50 GeV/c for both $\pi$ and $K$, corresponding 
to 1.5$\sigma$ mass separation between the two hypothesis.

\begin{figure}[!htb]
\includegraphics[width=0.45\textwidth]{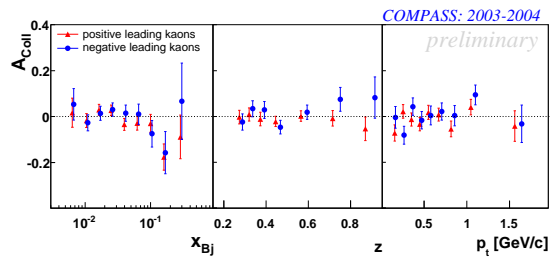}
\caption{Collins asymmetries for positive (triangles) and negative (circles) leading K as a function of $x$, $z$ and $p_T$.\label{fig:col_k}}

\end{figure}

\begin{figure}[!htb]
\includegraphics[width=0.45\textwidth]{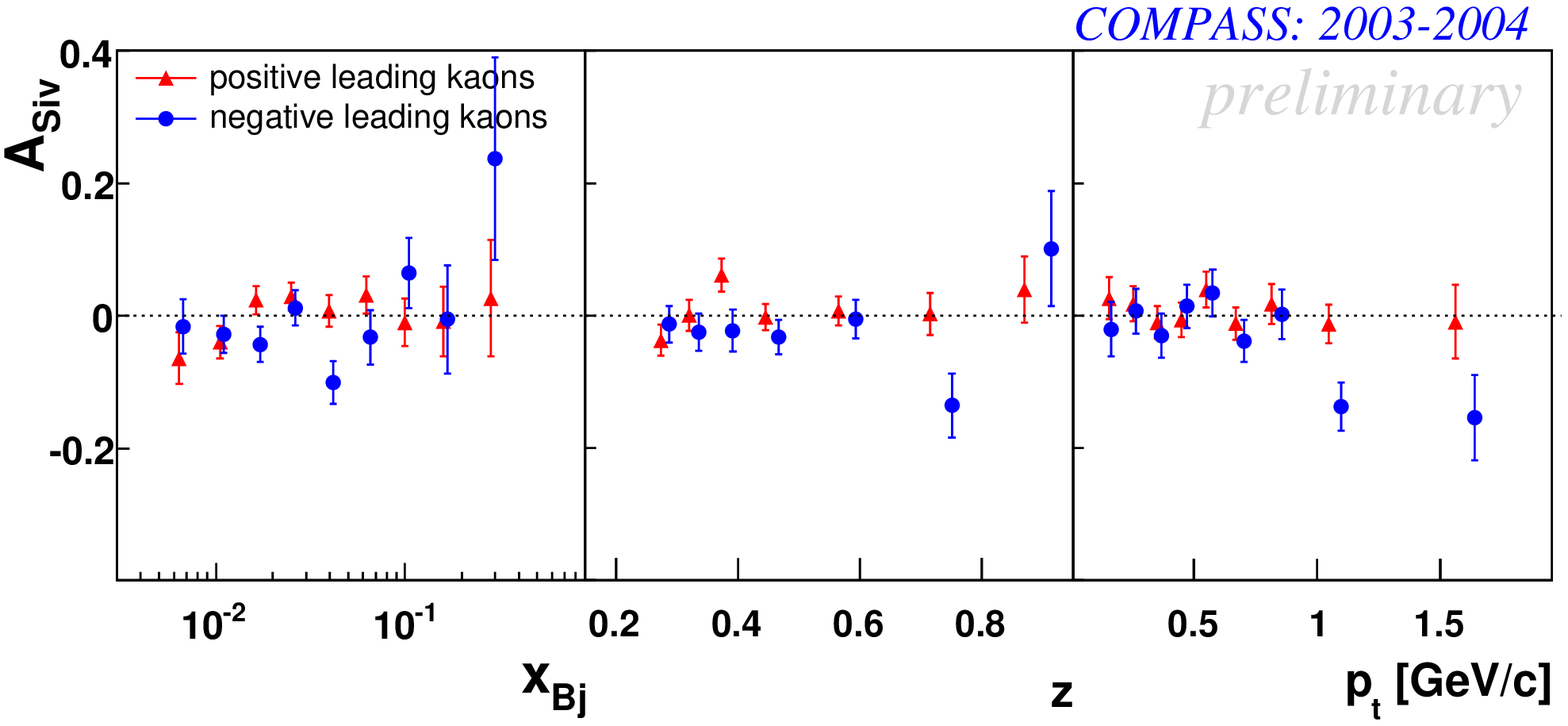}
\caption{Sivers asymmetries for positive (triangles) and negative (circles) leading K as a function of $x$, $z$ and $p_T$.\label{fig:siv_k}}

\end{figure}

After the identification,
 the final hadron sample consist of $5.2\cdot 10^6$ and $4.5 \cdot 10^6$ positive and negative $\pi$,
and $0.9\cdot 10^6$ and $0.45 \cdot 10^6$ positive and negative $K$.
The results on pion sample are very similar to those on unidentified hadrons (since more than 80\% of the hadrons are pions). The results on K sample are shown in fig.~\ref{fig:col_k} and~\ref{fig:siv_k}; also in this case, the asymmetries  are very small and compatible with zero within the statistical errors.

\section{HADRON PAIR ASYMMETRIES}
A different probe of the transversity function is the hadron pair production. 

The fragmentation function of a quark into a pair of charged hadrons is expected to exhibit an azimuthal dependence~\cite{coll_heppe}-\cite{bacchetta}, so that the number of events can be written as:

 \begin{equation}
N = N_0 \cdot (1+f  \cdot P_t  \cdot D_{nn}\cdot A_{\phi_{RS}}  \cdot \sin(\phi_{RS})) 
\end{equation}
where $\phi_{RS}$ is defined as $\phi_{R}-\phi_{s'}$. The $\phi_{R}$ angle is the angle between the lepton scattering plane and the plane containing the virtual photon momentum \textbf{q} and the component $R_T$ of the relative hadron momentum \textbf{R}$=\frac{1}{2}$(\textbf{P$_1$}-\textbf{P$_2$}) which is perpendicular to the summed hadron momentum \textbf{P$_h$}$=$\textbf{P$_1$}+\textbf{P$_2$}. 
The asymmetry $A_{\phi_{RS}}$ is 
proportional to the transversity function convoluted with the fragmentation function describing the two hadrons production, $H_1^{\angle h}$:
\begin{equation}
A_{\phi_{RS}} \propto \frac{\sum_q e_q^2 \, \Delta_T q(x) \, H_1^{\angle h}(z, M_h^2)}       {\sum_q e_q^2 \,     q(x) \,       D_q^h(z, M_h^2)} 
\end{equation}
where $M_h^2$ is the invariant mass of the hadron pair and $z=z_1+z_2$. 

\begin{figure}[htb]
\includegraphics[width=0.49\textwidth]{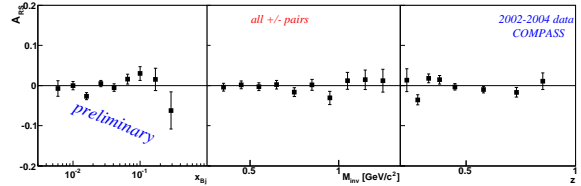}
\caption{ Asymmetries $A_{\phi_{RS}}$ for charged hadron pairs  as a function of $x$,  $M_h$ and $z$.}
\label{fig:2h}
\end{figure}

The   asymmetries have been measured on all the charged hadron pairs, 
for a total of $\sim$6.1$\cdot $10$^6$ combinations. 
 The results as a function of $x$, $M_h$ and $z$,
 are presented in fig.~\ref{fig:2h}. The errors shown are only the statistical contribution.
 The measured asymmetries are compatible with zero within the errors also in this case. 

\section{CONCLUSIONS}

COMPASS is the first experiment probing transversity in DIS on a deuterium target. 
Our measured asymmetries are small and compatible with zero, within the measured errors,
which typically are of the order of 1\%.
Since the HERMES collaboration 
have measured non zero asymmetries using  a proton target, the most probable interpretation 
for our data is a cancellation between 
the contribution coming from the proton and the neutron. 

COMPASS is planning to have a transversity data taking with a proton target in 2007.

\end{document}